\documentclass[12pt]{article}
\usepackage{color, amsmath, amsthm}

\usepackage{graphicx}
\usepackage{amsfonts}
\usepackage{amssymb,amsmath,amscd}
\usepackage[top=2.75cm, bottom=2.75cm, left=2.75cm, right=2.75cm]{geometry} 

\newtheorem{proposition}{Proposition}

\newtheorem{remark}[proposition]{Remark}

\def\+{{+\!\!\!+}}

\def\d{\partial}

\def\pmb#1{\setbox0=\hbox{#1}%
\kern.0em\copy0\kern-\wd0 
\kern-.04em\copy0\kern-\wd0 
\kern.08em\copy0\kern-\wd0 
\kern-.04em\raise.0433em\box0 }         
 
\newcommand{\nc}{\newcommand} 
\nc{\beq}{\begin{equation}} 
\nc{\eeq}[1]{\label{#1}\end{equation}} 
\nc{\ber}{\begin{eqnarray}} 
\nc{\eer}[1]{\label{#1}\end{eqnarray}} 
\nc{\pek}[1]{\cite{#1}} 
\nc{\enr}[1]{(\ref{#1})} 
\nc{\kal}[1]{{\cal{#1}}} 
\nc{\dott}{\;\cdot\;} 
\nc{\coker}{\mathrm{coker}}
\nc{\ie}{{\it i.e.}}
\nc{\eg}{{\it e.g.}}
\newcommand{\Section}[1]{\section{#1} \setcounter{equation}{0}} 

\numberwithin{equation}{section}

\def\0 {\nonumber}

\begin{document} 
\setcounter{page}{0}
\newcommand{\inv}[1]{{#1}^{-1}} 
\renewcommand{\theequation}{\thesection.\arabic{equation}} 
\newcommand{\be}{\begin{equation}} 
\newcommand{\ee}{\end{equation}} 
\newcommand{\bea}{\begin{eqnarray}} 
\newcommand{\eea}{\end{eqnarray}} 
\newcommand{\re}[1]{(\ref{#1})} 
\newcommand{\qv}{\quad ,} 
\newcommand{\qp}{\quad .} 

\thispagestyle{empty}
\begin{flushright} \small
UUITP-15/09  \\
NSF-KITP-09-57
 \end{flushright}
\smallskip
\begin{center} \LARGE
{\bf  Non-linear sigma models via\\
the chiral de Rham complex}
 \\[12mm] \normalsize
{\bf Joel~Ekstrand$^a$, Reimundo~Heluani$^b$, \\
Johan~K\"all\'en$^a$ and Maxim Zabzine$^a$} \\[8mm]
 {\small\it
 $^a$Department of Physics and Astronomy, 
     Uppsala university,\\
     Box 516, 
     SE-75120 Uppsala,
     Sweden\\
     ~\\
     $^b$Department of Mathematics, University of California,\\
      Berkeley, CA 94720, USA
}
\end{center}
\vspace{7mm}
\begin{abstract}
 \noindent  
  We propose a physical interpretation of the chiral de Rham complex as a formal Hamiltonian 
   quantization of the supersymmetric non-linear sigma model.   
   We show that the chiral de Rham complex
    on a Calabi-Yau manifold carries all information about the classical dynamics of the sigma model.
          Physically,  this provides an operator realization of  the non-linear sigma model.
     Mathematically, the idea suggests the use of Hamiltonian flow equations within the vertex algebra formalism 
      with the possibility to incorporate both left and right moving sectors within one mathematical framework. 
\end{abstract}

\eject
\normalsize



\section{Introduction}
\label{start}

The chiral de Rham complex (CDR) is a notion introduced by the mathematicians Malikov, Schechtman and Vaintrob
in \cite{Malikov:1998dw}.  CDR is  a sheaf of supersymmetric vertex algebras over a smooth manifold $M$. 
  It is defined by gluing free chiral algebras  on the overlaps of open subsets of $M$ isomorphic to $\mathbb{R}^n$
   ($\mathbb{C}^n$).   Since the original article  \cite{Malikov:1998dw},
   there has been considerable progress 
    on the mathematical literature about CDR.
In physics, however, CDR did not attract extensive attention. The key word in the physics interpretation 
   of CDR were the words ``chiral'' and ``perturbative''.  
  In  \cite{Kapustin:2005pt, Witten:2005px, Tan:2006qt, Tan:2007bh}  CDR (and more generally, chiral differential operators)
    were interpreted in the context of 
   the half-twisted sigma model in the perturbative regime.  In
   \cite{Malikov2006} a Lagrangian approach to CDR and other related
   sheaves was described. In another set of related works,
    \cite{Frenkel:2005ku, Frenkel:2006fy, Frenkel:2008vz}, CDR was discussed in the context of 
      the infinite volume limit of the sigma model. The infinite volume sigma
      model was suggested as a ``non-perturbative
       completion of CDR''. 

 In this work, we would like to initiate a different interpretation of CDR.  
Our idea originates from the observation that many formulas in two seemingly unrelated 
 subjects  are identical. Namely, the formulas originating from the  study of CDR \cite{Heluani1, Heluani2, Heluani:2008hw}
  are identical (modulo some quantum terms) to the formulas arising within the  classical Hamiltonian analysis
   of the $N=(1,1)$ supersymmetric non-linear  sigma model \cite{Zabzine:2005qf, Bredthauer:2006hf, Zabzine:2006uz}.
  Moreover, the Poisson brackets in the Hamiltonian formalism agree up to quantum terms with the quantum brackets 
   in CDR. This fact strongly suggests to interpret CDR as a formal canonical quantization of the non-linear sigma model. 
     However, CDR is just a formal quantization of the string phase space, and it does not carry any dynamical information
       unless the Hamiltonian flow equations are introduced in the game. If this is done, then CDR, with a particular 
        choice of a global section, encodes everything about the classical dynamics of the non-linear sigma model. Hopefully, 
         it knows about the quantum dynamics as well.  
       
 One peculiarity of the Hamiltonian formalism is that it mixes the left (chiral) and right (anti-chiral) moving sectors. 
  Thus, taking seriously the Hamiltonian interpretation of CDR, the word ``chiral''  in CDR becomes misleading. 
   If properly interpreted, CDR carries information about both the chiral
   and the anti-chiral sector of theory\footnote{The name
   chiral-anti-chiral de Rham complex was suggested in
   \cite{Frenkel:2008vz}.}.  
 Mathematically, it gives hope to recover the chiral and anti-chiral sectors within the framework of vertex algebras
  and therefore it may provide, eventually, a more solid mathematical basis for the study of non-linear sigma models. 

The paper is organized as follows. In section \ref{review} we briefly review the formalism of vertex algebras and 
the definition of CDR as a family of sheaves of  SUSY vertex algebras.
  In section \ref{CY} we discuss a particular set of global sections of CDR on a Calabi-Yau manifold which 
   gives rise to two commuting copies of the $N=2$ superconformal algebra. 
   Section \ref{sigma} reviews some basics about the classical $N=(1,1)$ supersymmetric sigma models and 
    its Hamiltonian treatment. We recover the above mentioned global sections of CDR within the Hamiltonian formalism. 
 We also discuss the classical equations of motion.  
  Section \ref{interpretation}   contains an interpretation of CDR as a formal 
   canonical quantization of the non-linear sigma model. We also show in
   this section how  CDR suggests the way to calculate 
    equal time commutators for the sigma model. 
     We also briefly discuss the mathematical aspects of the present interpretation.  In section  
     \ref{free} we present rather simple examples of interpreting the free boson and free fermion in terms of 
      CDR. 
Section \ref{end} contains the final remarks and further speculations on the present interpretation of CDR. 

Before proceeding further, let us make a disclaimer. This is a physics
article, not a mathematical one.  However, 
 we believe that  the present physical ideas bring along interesting
 mathematics and thus we allow
  ourselves for some short speculations on this subject. 
  We hope to return to proper mathematical treatment of these ideas in a separate publication. 

\section{Review of CDR}
\label{review}

 In this section we review the basics on vertex algebras and CDR. We also set the conventions for
   the rest of the article.  We use the name ``chiral de Rham complex'' (CDR) due to historical reasons. 
    Otherwise, we find the name CDR misleading, especially in the context of the present work.
     To avoid any fixation on the word ``chiral'' we denote 
     our formal coordinate for  the  punctured disk by $\xi$. In later discussions, we associate  $\xi = e^{i\sigma}$ with the 
      periodic  coordinate $\sigma$ along the loop.  CDR was originally introduced in \cite{Malikov:1998dw}, however, we follow the treatment given in \cite{Heluani1}. 

First, let us  review the definition of {\it vertex superalgebras},
as presented in \cite{Kac:1996wd}.
    Given a vector space $V$, an \emph{$\rm{End}(V)$-valued field} is a formal
    distribution of the form
    \begin{equation*}
        A(\xi) = \sum_{n \in \mathbb{Z}} \xi^{-1-n} A_{(n)},\qquad A_{(n)} \in
        \rm{End}(V),
    \end{equation*}
    such that for every $v \in V$, we have $A_{(n)}v = 0$ for large enough $n$.

    A vertex superalgebra consists of the data of a super vector space $V$,
    an even vector $|0\rangle \in V$ (the vacuum vector),
     an even endomorphism $\d$, and a parity preserving linear map $A \mapsto Y(A,\xi)$ from
     $V$ to $\rm{End}(V)$-valued fields (the state-field correspondence). This
     data should satisfy the following set of axioms:
    \begin{itemize}
    \item Vacuum axioms:
        \begin{equation*}
            \begin{aligned}
            Y(|0\rangle, \xi) &= \rm{Id}, \\
            Y(A, \xi) |0\rangle &= A + O(\xi), \\
            \d |0\rangle &= 0.
            \end{aligned}
        \end{equation*}
    \item Translation invariance\footnote{We denote the even endomorphism and the derivative along $\xi$ by the
     same sign $\d$. The appropriate interpretation of $\d$ should be clear from the context.}:
        \begin{equation*}
            \begin{aligned}
                {[}\d, Y(A,\xi)] &= \partial_\xi Y(A,\xi).
        \end{aligned}
        \end{equation*}
    \item Locality:
        \begin{equation*}
            (\xi-\xi')^n [Y(A,\xi), Y(B,\xi')] = 0, \qquad n \gg 0.
        \end{equation*}
     \end{itemize}
(The notation $O(\xi)$ denotes a power series in $\xi$ without constant
term.)
 
    Given a vertex super-algebra $V$ and a vector $A \in V$, we expand the fields
    \begin{equation*}
        Y(A,\xi) = A(\xi) =  \sum_{{j \in \mathbb{Z}}} \xi^{-1-j}
        A_{(j)},
    \end{equation*}
    and we call the endomorphisms $A_{(j)}$ the \emph{Fourier modes} of
    $Y(A,\xi)$. Now, define the operations
    \begin{equation*}
        \begin{aligned}
            {[}A_\lambda B] &= \sum_{{j \geq 0}}
            \frac{\lambda^{j}}{j!} A_{(j)}B, \\
            :A B: &= A_{(-1)}B,
        \end{aligned}
    \end{equation*}
     where $\lambda$ is a formal even parameter. 
    The first operation is called the $\lambda$-bracket and the second is
    called the \emph{normally ordered product}.
     The $\lambda$-bracket contains all the information about the commutators between the Fourier 
      coefficients of fields in $V$, and the OPE can easily  be read from it, namely
      \beq
       A(\xi) B(\xi') = \sum\limits_{j \geq 0}  \frac{ \Bigl(A_{(j)}
       B\Bigr)(\xi')}{(\xi- \xi')^{j+1}} + : A(\xi) B(\xi'):~,
      \eeq{opefromlambda}
       where the last term corresponds to the normally ordered product.
       Correspondingly, the commutator is 
       \beq
        [A(\xi), B(\xi')] = \sum\limits_{j \geq 0}  \frac{1}{j!}
	~(A_{(j)} B)(\xi')~\d^j_{\xi'} \delta (\xi-\xi')~.
       \eeq{commutatorssw93}
       In all further considerations we drop the notation 
        $:~:$, and the normal ordering is always assumed in the quantum
	setup.  In practice, the $\lambda$-bracket is a convenient 
        tool for manipulations with OPEs.

  Next, we review {\it SUSY vertex algebras} from \cite{heluani3}. The idea behind the definition is to extend
   the formal even variable $\xi$ to  formal variables $(\xi,\theta)$
   with $\theta$ being an odd Grassman variable. 
     Given a super vector space $V$, an $\rm{End}(V)$-valued superfield is a
     formal distribution of the form
    \begin{equation*}
       A(\xi, \theta)=   \sum_{\stackrel{j \in \mathbb{Z}}{J = 0,1}} \xi^{-1-j} \theta^{1-J} A_{(j|J)}~,
        \qquad A_{(j|J)} \in \rm{End}(V)~,
    \end{equation*}
     such that for every $v\in V$, $A_{(j|J)} v =0$ for large enough $j$.
     
 A SUSY vertex algebra consists of the data of a super vector space $V$,
    an even vector $|0\rangle \in V$ (the vacuum vector), an odd endomorphism
    $D$ (whose square is an even endomorphism which we denote $\d$),
    and a parity preserving linear map $A \mapsto Y(A,\xi, \theta)$ from
     $V$ to $\rm{End}(V)$-valued superfields (the state-superfield correspondence). This
     data should satisfy the following set of axioms:
    \begin{itemize}
    \item Vacuum axioms:
        \begin{equation*}
            \begin{aligned}
            Y(|0\rangle, \xi, \theta) &= \rm{Id}~, \\
            Y(A, \xi, \theta) |0\rangle &= A + O(\xi, \theta)~, \\
            D |0\rangle &= 0~.
            \end{aligned}
        \end{equation*}
    \item Translation invariance:
        \begin{equation*}
            \begin{aligned}
            {[} D, Y(A,\xi, \theta)] &= (\partial_\theta - \theta \partial_\xi)
            Y(A,\xi, \theta),\\
            {[}\d, Y(A,\xi, \theta)] &= \partial_\xi Y(A,\xi, \theta).
        \end{aligned}
        \end{equation*}
    \item Locality:
        \begin{equation*}
            (\xi-\xi')^n [Y(A,\xi, \theta), Y(B,\xi', \theta')] = 0
	    \qquad n \gg 0.
        \end{equation*}
     \end{itemize}
    \label{defn:2.3}
 (The notation $O(\xi, \theta)$ denotes a power series in $\xi$ and
 $\theta$ without a constant
term in $\xi$.)
 
    Given the vacuum axioms for a SUSY vertex algebra, we will use the state-field correspondence to identify a vector $A \in V$ with its corresponding
    field $Y(A,\xi, \theta)$.
    \label{rem:nosenosenose}
    Given a SUSY vertex algebra $V$ and a vector $A \in V$, we expand the fields
    \begin{equation*}
        Y(A,\xi, \theta) = A(\xi, \theta)= \sum_{\stackrel{j \in \mathbb{Z}}{J = 0,1}} \xi^{-1-j} \theta^{1-J}
        A_{(j|J)},
    \end{equation*}
    and we call the endomorphisms $A_{(j|J)}$ the \emph{Fourier modes} of
    $Y(A,\xi, \theta)$. Define now the operations
    \begin{equation}
        \begin{aligned}
            {[}A_\Lambda B] &= \sum_{\stackrel{j \geq 0}{J = 0,1}}
            \frac{\Lambda^{j|J}}{j!} A_{(j|J)}B, \\
            A B &= A_{(-1|1)}B,
        \end{aligned}
        \label{eq:2.4.2}
    \end{equation}
     where $\Lambda^{j|J} = \lambda^j \chi^J$, with formal even $\lambda$ and odd $\chi$ satisfying 
       $\chi^2 = - \lambda$. 
    The first operation is called the $\Lambda$-bracket and the second is
    called the \emph{normally ordered product}. This $\Lambda$-bracket is
    an efficient way to encode and manipulate
      OPEs of superfields.  In particular, the commutator of two
      superfields is given
      by the following expression:
     \beq
      [A(\xi, \theta), B(\xi', \theta')] = \sum_{\stackrel{j \geq 0}{J =
      0,1}} \frac{(-1)^J}{j!}     \Bigl(\d_{\xi'}^j D_{\xi'\theta'}^J
      \delta(\xi-\xi')\delta(\theta-\theta')\Bigr)~(A_{(j|J)}B) (\xi', \theta') ~. 
       \eeq{slsl292020aaa} 
     As in the standard setting, given a SUSY vertex algebra $V$ and a vector $A \in V$, we
    have:
    \begin{equation*}
        Y(\d A, \xi, \theta) = \partial_\xi Y(A,\xi, \theta) = [\d, Y(A, \xi, \theta)]~.
    \end{equation*}
    On the other hand, the action of the derivation $D$ is described
    by:
    \begin{equation*}
        Y(DA,\xi, \theta) = \left( \partial_\theta + \theta \partial_\xi \right) Y(A,\xi, \theta)
        \neq [D, Y(A,\xi, \theta)].
    \end{equation*}
    \label{rem:caca5}
 For further details of the formalism the reader may consult \cite{heluani3}. 

In \cite{Malikov:1998dw}, given any smooth manifold $M$, 
the authors introduced a sheaf of vertex algebras on $M$ which they 
 called the chiral de Rham complex (CDR). Roughly, the idea is  to associate
 locally over a neighborhood of a point, a 
 vertex algebra corresponding to a free $\beta\gamma bc$-system.  The crucial observation is that the group of coordinate 
  changes can be mapped into the group of vertex algebra  automorphisms
  of this free system.  This allows one to glue the algebras associated
  to different open sets in $M$ together, and 
   construct a sheaf.  Although the formalism of \cite{Malikov:1998dw}
   works in the analytic, algebraic and smooth settings, most of the
   mathematics literature on CDR is dedicated to the algebraic case.  In
   the present case, we are interested 
    in the smooth setting. This set-up was considered in \cite{lian} and,
    in the superfield formulation, in \cite{Heluani1}. 

 The $\beta\gamma$-system is the vertex algebra generated by the even fields
 $\beta_\mu$ and $\gamma^\mu$, subject to the following relation:
 \beq
  [\beta_\mu (\xi), \gamma^\nu (\xi')] = \hbar \delta_\mu^\nu \delta(\xi-\xi')~.
 \eeq{betagannaregs}
The fermionic $bc$-system is the vertex algebra generated by the odd
fields $b_\mu$ and $c^\mu$ satisfying
\beq
[c^\mu(\xi), b_\nu(\xi')]_+ = \hbar \delta^\mu_\nu \delta (\xi- \xi')~,
\eeq{betahaahhsbc}
 where $[~,~]_+$ stands for anticommutator. In the following discussion,
 all our operations will be $\mathbb{Z}_2$-graded,
  and we will drop the subscript $+$.  In section \ref{free} we will consider
  the $\beta\gamma bc$-system in 
   more detail.  We can combine these fields into superfields 
and introduce the SUSY vertex algebra generated by 
\beq
 \phi^\mu (\xi, \theta) = \gamma^\mu(\xi) + \theta c^\mu(\xi)~, \quad \quad \quad
  S_\mu (\xi, \theta) = b_\mu(\xi) + \theta \beta_{\mu}(\xi)~,
\eeq{defhfjj222}
which satisfy 
 \beq
   [\phi^\mu (\xi, \theta),  S_\nu(\xi', \theta') ] = \hbar \delta^\mu_\nu \delta (\xi- \xi') \delta (\theta-\theta')~,
 \eeq{jdjdkkww}
 or equivalently:
$$ {[\phi^\mu}_\Lambda S_\nu] = \hbar \delta^\mu_\nu ~. $$
  From (\ref{jdjdkkww}) we easily recover the collection of the standard $\beta\gamma$-system (\ref{betagannaregs})  
   and $bc$-system (\ref{betahaahhsbc}).
   Given a change of coordinates $\tilde{x}^\mu = g^\mu(x)$ with its
   inverse $x^\mu = f^\mu (\tilde{x})$, we can 
    define a SUSY vertex algebra automorphism as follows:
\beq
 \tilde{\phi}^\mu = g^\mu (\phi)~,~~~~~~~~~~~~\tilde{S}_\mu= \left (
 \frac{\d f^\mu}{\d \tilde{\phi}^\nu} (g (\phi)) S_\nu \right)~.
\eeq {quantumtragshjj}
Recall that products are normally ordered.
The automorphism (\ref{quantumtragshjj}) preserves, in particular, the relation
(\ref{jdjdkkww}).  Using (\ref{quantumtragshjj}) we can glue the SUSY
vertex algebras generated by (\ref{jdjdkkww})
 to obtain a sheaf  $\Omega^{\mathrm{ch}}_\hbar (M)$ of SUSY vertex algebras over
 $M$.  Indeed, we deal with a family of 
  sheaves which depends (polynomially) on $\hbar$.
   It is important to stress that at this moment we do not need to discuss any allocation of 
     conformal weights for the fields $\phi^\mu$ and $S_\mu$. The set of
     global sections $\Gamma(M,\Omega^{\mathrm{ch}}_\hbar (M))$
     give rise to a family (depending on $\hbar$) of SUSY vertex algebras 
     attached to $M$.

\section{CDR on Calabi-Yau manifolds}
\label{CY}
 In this section, we construct a specific collection of global sections of
 $\Omega^{\mathrm{ch}}_\hbar (M)$ on any Calabi-Yau manifold. 
  In our presentation, we borrow the results from \cite{Heluani1,
  Heluani2}.  
  
 On any orientable manifold $M$, with the choice of volume form
   $${\rm vol}_d = e^{\rho(x)}~ dx^1\wedge ... \wedge dx^d~,$$
  the following defines a global section
 of $\Omega^{\mathrm{ch}}_\hbar (M)$:
\beq
 {\cal P} = D\phi^\mu DS_\mu + \d \phi^\mu S_\mu - \hbar \d D \rho~.
\eeq{slslH}
Expanded as 
 $${\cal P}(\xi, \theta) = G(\xi) + 2\theta L(\xi)~,$$
  it generates the $N=1$ superconformal algebra $(G, L)$ with central charge $3\dim M$. 
  With respect to this Virasoro generator $L$, the fields $\phi^\mu$ are primary
  of conformal weight $0$. The fields $S_\mu$ have conformal weight $1/2$, but
  are not primary unless
  we choose coordinates where the volume form is constant. We stress that
  this assignment of conformal weights is not the one that usually is
  considered in the literature, in particular, it differs from the one in
  the original work \cite{Malikov:1998dw}. For further details, see
  section \ref{twist} below. 

 Let $M$ be a complex manifold with complex structure $I$, and suppose  
  it admits a closed holomorphic volume form $\Omega$:
  $$\Omega = e^{f(z)} dz^1 \wedge ... \wedge dz^n~,$$
   written in holomorphic coordinates, with $\dim_{\mathbb R} M = d =
   2n$. $\Omega$ is related to the real volume form by
    $$  \Omega \wedge \bar{\Omega} = i^{n(n+2)} 2^n ~{\rm vol}_d~.$$ 
In this case, $\Omega^{\mathrm{ch}}_\hbar (M)$ has the following global
 section: 
\beq
{\cal J}_1 = (I^\mu_\nu D \phi^\nu) S_\mu  + \frac{i}{2}  \hbar D(f-\bar{f}) ~,
\eeq{3399skskkk}
such that 
\ber
&& {\cal J}_1(\xi, \theta) = -i J_1(\xi) - i \theta (G_1^-(\xi) - G_1^+(\xi) )~,\\
 && {\cal P} (\xi, \theta) = (G_1^+(\xi) + G_1^-(\xi)) + 2\theta L_1(\xi)~, 
\eer{djdjks9393}
 and $(J_1, G_1^\pm, L_1)$ generate the $N=2$ superconformal algebra of central charge $3 \dim M$. 

 If the manifold is symplectic with the symplectic form $\omega$, 
  then CDR admits the following global section:
 \beq
 {\cal J}_2 = \frac{1}{2} \left ( \omega^{\mu\nu} S_\mu S_\nu -\omega_{\mu\nu} D \phi^\mu D \phi^\nu  \right )~,
\eeq{skskwow999}
 with $\omega_{\mu\nu} \omega^{\nu\lambda} = \delta_\mu^\lambda$, such that 
  \ber
&& {\cal J}_2(\xi, \theta) = -i J_2(\xi) - i \theta (G_2^-(\xi) - G_2^+(\xi) )~,\\
 && {\cal P} (\xi, \theta) = (G_2^+(\xi) + G_2^-(\xi)) + 2\theta L_2(\xi)~, 
\eer{djdjks9393a}
 and $(J_2, G_2^\pm, L_2)$ generate the $N=2$ superconformal algebra of central charge $3 \dim M$. 
  The existence of the global sections (\ref{3399skskkk}) and (\ref{skskwow999})
  and their relation to $N=2$ superconformal algebra  is a generic feature of generalized Calabi-Yau manifolds
   \cite{Heluani:2008hw}. 
   
   Let $M$ be a Calabi-Yau manifold and choose  a Ricci  flat
  metric $g$, a complex structure $I$ and 
   a closed K\"ahler form $\omega = gI$. In this case 
     $\rho = \log \sqrt{\det (g_{\mu\nu})}$ and, in addition to the
     sections (\ref{slslH}), (\ref{3399skskkk}) and (\ref{skskwow999}),
    one can define another global section\footnote{${\cal H}$ is a global
    section on a Calabi-Yau manifold, see \cite{Heluani2} for further 
     details.  It is not clear whether or not it is a global section on a generic Riemannian manifold.} 
\beq
 {\cal H} = \d \phi^\mu D\phi^\nu g_{\mu\nu} + g^{\mu\nu}  DS_\mu S_\nu
 +     \Gamma^\rho_{~\sigma\nu}   g^{\nu\lambda} D\phi^\sigma  (S_\lambda S_\rho )~,
\eeq{H'}
where $\Gamma$ is the Levi-Civita connection.  The algebraic relations
satisfied by these global sections $({\cal P}, {\cal J}_1, {\cal J}_2, {\cal H})$ are
 given in the appendix, both in terms of $\Lambda$-brackets:
 (\ref{LLLL1})-(\ref{LLLL10}), and of their commutators:
 (\ref{askssl111})-(\ref{2202wkkwk2h2}).
 Furthermore,  we define the following global sections:
\beq
   {\cal H}_L = \frac{1}{2} ({\cal P} + {\cal H})~,~~~~~~{\cal H}_R= \frac{1}{2} ({\cal P} - {\cal H})~,~~~~~~
   {\cal J}_L = \frac{1}{2} ({\cal J}_1 + {\cal J}_2)~,~~~~~~ {\cal J}_R = \frac{1}{2} ({\cal J}_1 - {\cal J}_2)~,
\eeq{twocopiesss}
  such that 
\ber
&& {\cal J}_L(\xi, \theta) = - i J_L (\xi) - i \theta (G_L^-(\xi) - G_L^+(\xi))~,\\
 &&{\cal H}_L (\xi, \theta) = (G_L^+(\xi) + G_L^-(\xi)) + 2\theta L_L (\xi)~,\\
 && {\cal J}_R (\xi, \theta) = - i J_R (\xi) - i \theta (G_R^-(\xi) - G_R^+(\xi))~,\\
 && {\cal H}_R (\xi, \theta) = (G_R^+(\xi) + G_R^-(\xi)) + 2\theta L_R
 (\xi)~.
\eer{ddkk303003idi}
 A tedious calculation from \cite{Heluani2} shows that $(J_L, G_L^\pm, L_L)$ and $(J_R, G_R^\pm, L_R)$ generate 
  two commuting copies of the $N=2$ superconformal algebra with central charge $\frac{3}{2} \dim M$ each.  The relation between 
   the different $N=2$ algebras is given by the following expressions:
\ber
\label{AAHH1}&& L=L_1=L_2 = L_L + L_R~,\\
\label{AAHH2} && J_1 = J_L + J_R~,~~~~J_2 = J_L- J_R~,\\
 \label{AAHH3}&& G_1^\pm = G_L^\pm + G_R^\pm~,~~~~G_2^\pm = G_L^\pm + G^\mp_R~,\\
\label{AAHH4} && G = G_1^+ + G_1^- = G_2^+ + G_2^-~. 
\eer{sjsjeieie9399393}
Summarizing, CDR enables us to construct four different $N=2$
superconformal algebras on a Calabi-Yau manifold with fixed complex and
K\"ahler moduli. 
 There are two commuting copies   $(J_L, G_L^\pm, L_L)$ and $(J_R, G_R^\pm, L_R)$,
  with central charge $\frac{3}{2} \dim M$ each. Their
  different ``diagonal'' combinations 
  (\ref{AAHH1})-(\ref{AAHH4}) give rise to the $N=2$ algebra $(J_1,
  G_1^\pm, L_1)$ which depends only on the complex moduli;
   and to the $N=2$ algebra  $(J_2, G_2^\pm, L_2)$ which depends only on the K\"ahler moduli. 

The following is a simple but important observation about these two
commuting copies of the $N=2$ superconformal algebra. 
 Let us map the variable $\xi$ to a loop coordinate $\sigma$ such that $\xi = e^{i\sigma}$.
  The transformations will be controlled by the diagonal Virasoro algebra $L$, see (\ref{AAHH1}). 
   Thus a field $O(\xi)$, primary of conformal weight $\Delta$:
   $$ O(\xi) = \sum\limits_{n\in {-\Delta + \mathbb Z}} O_n \xi^{-n - \Delta},$$
   has the following expansion with respect to $\sigma$:
  $$ O(\sigma) =  i^{\Delta}\sum\limits_{n \in {\mathbb Z}} O_n e^{-in\sigma}~.$$
  We may regard these operators as operators at time zero within the canonical quantization 
   on a cylinder $S^1  \times \mathbb{R}$. 
 We choose
  the Hamiltonian 
\beq
 H = i (L_L)_0 - i (L_R)_0 = \frac{i}{2} \int d\xi d\theta \xi~ {\cal
 H}(\xi, \theta)~,
\eeq{sksksk229}
and postulate the following flow equation for an operator $O$:
\beq
 \frac{d O(\sigma, t)}{d t} =\frac{1}{\hbar} [H, O(\sigma, t)]~.
\eeq{floweqs}
  The formal solution to the flow
  equation (\ref{floweqs}) is  
   \beq
  O(\sigma, t) = e^{\frac{1}{\hbar}t H} O(\sigma ) e^{-\frac{1}{\hbar}t  H}~.
 \eeq{flowsolution}
  This setup allows us to interpret the model as defined over the cylinder. 
  In particular, we get from (\ref{sksksk229}) and (\ref{floweqs})
 that $J_L(t + \sigma), G_L^\pm (t + \sigma), L_L(t+\sigma)$ and $J_R(t - \sigma), G_R^\pm (t - \sigma), 
  L_R(t-\sigma)$. Thus we have left and right moving $N=2$ algebras defined over a cylinder.

\section{Classical sigma model}
\label{sigma}
 In this section we consider the classical  supersymmetric non-linear sigma model defined over $S^1 \times {\mathbb R}$, 
  with Minkowski signature. The Euclidean case can be treated along the same lines.
  
The supersymmetric $N=(1,1)$ sigma model is defined by the following action functional
\beq
 S = \frac{1}{2} \int d \sigma\, dt\, d\theta^- d\theta^+ D_+ \Phi^\mu D_-\Phi^\nu g_{\mu\nu}(\Phi)~,
\eeq{classical-action}
 where we use the $N=(1,1)$ superfield formalism and $g$ is a metric on the target $M$.
   The even coordinate $\sigma$ parametrizes the circle $S^1$ and 
  $t$ is the time coordinate for $\mathbb{R}$.  The pair $\theta^\pm$ labels the spinor coordinates. The spinor 
   derivatives $D_\pm$ are defined as:
\beq
 D_\pm = \frac{\d}{\d \theta^\pm} + \theta^\pm (\d_0 \pm \d_1)~,~~~~~~~~
 D^2_\pm = \d_0 \pm \d_1~,
\eeq{definderi33}
 where $\d_0 \equiv \frac{\d}{\d t}$ and $\d_1 \equiv \frac{\d}{\d
 \sigma}$. This action is invariant under 
  $N=(1,1)$ superconformal transformations. If the target manifold $M$ is K\"ahler then the action (\ref{classical-action}) 
   is invariant under $N=(2,2)$ superconformal symmetry and in particular
   it is invariant under the following transformations:
\beq
 \delta \Phi^\mu = \epsilon^+ D_+ \Phi^\nu I^\mu_\nu (\Phi) + \epsilon^- D_- \Phi^\nu I^\mu_\nu (\Phi)~,
\eeq{susyextraappp}
 where $\epsilon^\pm$ are odd functions subject to the condition $D_\pm
 \epsilon^\mp=0$, and $I$ is the complex structure on 
  the target $M$. 

We would like to define the Hamiltonian formalism for the model and discuss the Hamiltonian realization of the 
 symmetries of the action (\ref{classical-action}).  In going to Hamiltonian formalism we would like to get rid of 
  one odd $\theta$.  Let us introduce new odd coordinates as follows:
  \beq
  \theta^+ = \frac{1}{\sqrt{2}} (\theta_0 + \theta_1)~,~~~~~~~\theta^- =\frac{1}{i\sqrt{2}}(\theta_0 - \theta_1)~,
 \eeq{newoddcoordns}
  together with odd derivatives
 \beq
   D_+ = \frac{1}{\sqrt{2}} (D_0 + D_1)~,~~~~~~~~
   D_- = \frac{1}{i\sqrt{2}} (D_0 - D_1)~,
 \eeq{redefinsoowow}
 which satisfy  $D^2_0 = \d_1$, $D_1^2 = \d_1$ and 
  $D_1 D_0 + D_0 D_1 = 2 \d_0$. In order to integrate out $\theta_0$ we
  introduce  new superfields:
 \beq
  \phi^\mu = \Phi^\mu|_{\theta_0=0}~,~~~~~~~~~~~
  S_\mu = g_{\mu\nu} D_0 \Phi^\nu |_{\theta_0=0}~,
 \eeq{ddjdjj33999}
  and from now on $D_1 = D_1|_{\theta_0=0}$. After performing
  $\theta_0$-integration, 
   the action (\ref{classical-action}) becomes 
 \beq
 S= \int dt d\sigma d\theta_1 \left ( S_\mu \d_0 \phi^\mu - \frac{1}{2} {\cal H} \right )~,
\eeq{actionhamiltonian}
 where
\beq
 {\cal H} =  \d_1 \phi^\mu D_1 \phi^\nu g_{\mu\nu} + g^{\mu\nu} S_\mu D_1 S_\nu
 + S_\rho D_1 \phi^\gamma S_\lambda g^{\nu\lambda} \Gamma^\rho_{~\gamma\nu}~.
\eeq{hamiltoaniana}
 Thus we can conclude that the sigma model phase space  
  corresponds to  a cotangent bundle $T^* {\cal L}M$ to a superloop space ${\cal L}M= \{S^{1|1} \rightarrow M\}$
   equipped with the natural symplectic structure
\beq
  \int d\sigma\, d\theta_1 ~\delta S_\mu \wedge \delta \phi^\mu~. 
\eeq{syemsllspddp}
 Here $\theta_1$ transforms as a section of the square root of the canonical bundle over $S^1$. 
 Thus the space of  functionals on  $T^* {\cal L}M$ is equipped with
 a super-Poisson bracket $\{~,~\}$ generated 
 by the relation:  
   \beq
   \{ \phi^\mu (\sigma, \theta_1), S_\nu (\sigma', \theta'_1) \} =
   \delta^\mu_\nu \delta (\sigma - \sigma') \delta(\theta_1 -
     \theta'_1)~.
  \eeq{susyddhhdpp}
  From (\ref{actionhamiltonian}) and (\ref{hamiltoaniana})
    the Hamiltonian is:  
  $$H = \frac{1}{2} \int d\sigma d\theta_1 ~{\cal H},$$ 
   and the sigma model equation of motions are
 \begin{equation}
	 \begin{aligned}
   \frac{\d\phi^\mu}{\d t} &=  \{H,  \phi^\mu\} = g^{\mu\nu} D_1 S_\nu - 
   g^{\mu\sigma} \Gamma^\lambda_{~\sigma\nu} D_1 \phi^\nu
   S_\lambda~,\\
   \frac{\d S_\mu}{\d t} &= \{H, S_\mu\} =  g_{\mu\rho} \d_1 D_1 \phi^\rho +  \d_1 \phi^\nu D_1 \phi^\rho \Gamma_{\mu\nu\rho} +  S_\nu D_1 S_\rho g ^{\rho\lambda} \Gamma^\nu_{~\lambda\mu}  \\
      &\quad +  \frac{1}{2}S_\nu D_1 \phi^\rho S_\lambda \left  ( \d_\rho (g^{\lambda\gamma} \Gamma^\nu_{~\mu\gamma})       - \d_\mu (g^{\lambda\gamma} \Gamma^\nu_{~\rho\gamma}) \right )~,
  \label{ejdjw02020273}
  \end{aligned}
  \end{equation}
  where $\partial_\rho$ and $\partial_\mu$ in the last equation refer to
  derivatives along the target coordinates. The equations
  (\ref{ejdjw02020273})
   are  equivalent to the equations which follow from the variational principle for (\ref{classical-action}). 
 
 Next let us discuss how the symmetries of the action functional (\ref{classical-action}) are realized in the Hamiltonian 
  formalism. It is a tedious but straightforward computation to check
  that (\ref{susyextraappp}) are realized in the Hamiltonian 
   formalism  by the following generators:
  \beq
 {\cal J}_1 (\epsilon_1) = \int d\sigma d\theta_1 ~\epsilon_1 (\sigma, \theta_1) \, I^\mu_\nu D_1  \phi^\nu S_\mu  ~,
\eeq{hamsupercdnskk}
\beq
 {\cal J}_2(\epsilon_2) =  \frac{1}{2}  \int d\sigma d\theta_1 ~\epsilon_2  (\sigma, \theta)  \left ( \omega^{\mu\nu} S_\mu S_\nu 
  - \omega_{\mu\nu} D_1 
  \phi^\mu D_1 \phi^\nu \right )~,
\eeq{hamsupercdnskka}
 where $I$ is complex structure and $\omega = gI$ is K\"ahler form.  The
 parameters are related as follows:
  $$\epsilon_1 =\frac{1}{\sqrt{2}} ( i \epsilon_- + \epsilon_+  )~,~~~~~~~~~~~
  \epsilon_2 =  \frac{1}{\sqrt{2}} (   i \epsilon_- -\epsilon_+)~,$$
  where $\epsilon_\pm$ are evaluated at $t=0$. 
 The remaining generators of $N=(2,2)$ superconformal symmetry 
  can be calculated by computing the Poisson brackets between different 
  combinations of ${\cal J}_1$ and ${\cal J}_2$.  In addition to
  (\ref{hamsupercdnskk}) 
   and (\ref{hamsupercdnskka}) we obtain
  \beq
  {\cal P} (a_1) = \int d\sigma d\theta_1 ~  a_1(\sigma, \theta_1) \left ( D_1 \phi^\mu D_1 S_\mu 
  + \d_1  \phi^\mu S_\mu \right )~,
  \eeq{ssj33939931}
  and
  \beq
   {\cal H} (a_2)  = \int d\sigma d\theta_1~ a_2(\sigma, \theta_1) 
   \left ( \d_1 \phi^\mu D_1 \phi^\nu g_{\mu\nu} + g^{\mu\nu} S_\mu D_1 S_\nu
 + S_\rho D_1 \phi^\gamma S_\lambda g^{\nu\lambda} \Gamma^\rho_{~\gamma\nu} \right )~,
  \eeq{djjde00303eekekek}
 thus obtaining the Hamiltonian realization of the $N=(2,2)$ superconformal symmetries of
     the action (\ref{classical-action}). 

In order to compare these results with the expressions of the previous
section, we perform a change of coordinates. 
 Since we are in the classical set-up, we can easily change coordinates even at the Lagrangian level. 
  In particular, we perform the transformation $\xi = e^{i\sigma}$,
  $(i\xi)^{-1/2}\theta = \theta_1$, which
  imply:
  $$
 (i\xi)^{1/2}  D =  D_1~,~~~~~ (i\xi)^{1/2} d\theta =  d\theta_1~,~~~~~
 (i\xi) ^{1/2} S_\mu (\xi, \theta) = S_\mu (\sigma, \theta_1)~,$$
  where we now use the notations from the previous two sections. Now, one
  can easily see that all Hamiltonian 
   generators  of $N=(2,2)$ superconformal symmetry derived from the
   action principle are mapped directly to the 
    expressions from the previous section, modulo $\hbar$-terms,
    \ber
\label{FFFFF1}    {\cal P}(\xi, \theta) & = & D\phi^\mu DS_\mu + \d \phi^\mu S_\mu ~,\\
\label{FFFFF2}     {\cal J}_1(\xi, \theta) & = & (I^\mu_\nu D \phi^\nu) S_\mu ~, \\
 \label{FFFFF3}    {\cal J}_2 (\xi, \theta) &=  & \frac{1}{2} \left ( \omega^{\mu\nu} S_\mu S_\nu -\omega_{\mu\nu} D \phi^\mu D \phi^\nu  \right )~,\\
\label{FFFFF4}     {\cal H} (\xi, \theta) & =  &\d \phi^\mu D\phi^\nu g_{\mu\nu} + g^{\mu\nu} S_\mu DS_\nu
 + S_\rho D\phi^\sigma S_\lambda g^{\nu\lambda} \Gamma^\rho_{~\sigma\nu}~.
    \eer{djfkfkkeee}
      In particular the Hamiltonian becomes
 \beq
H = \frac{i}{2} \int d\xi d\theta\, \xi  \left ( \d \phi^\mu D\phi^\nu g_{\mu\nu} + g^{\mu\nu} S_\mu DS_\nu
 + S_\rho D\phi^\sigma S_\lambda g^{\nu\lambda} \Gamma^\rho_{~\sigma\nu}\right )~,
\eeq{hamiltoanian}
 which is the same as in (\ref{sksksk229}).  The conventions for the Poisson brackets and the brackets 
  for the above classical generators can be found in the appendix, see (\ref{bliablia282})-(\ref{2202wkkwk2h2ss}). 

\section{Interpretation}
\label{interpretation}

In this section we combine the results from the two previous sections. We offer both physical and mathematical
 interpretations of these results.

\subsection{Physical aspects}
\label{physics}
 In section \ref{sigma} we analyzed, in the Hamiltonian formalism,  
  the classical supersymmetric non-linear sigma model with a K\"ahler manifold as target. 
  Its  
  phase space is defined as the cotangent bundle to a superloop space  $T^* {\cal L}M$. The canonical Poisson 
   bracket is given in local coordinates by the following expression: 
 \beq
   \{ \phi^\mu (\xi, \theta), S_\nu (\xi', \theta') \} = \delta^\mu_\nu~ \delta (\xi - \xi') \delta (\theta - \theta')~. 
 \eeq{38383899} 
  Moreover, we have found the generators $({\cal P}, {\cal J}_1, {\cal J}_2, {\cal H})$ 
   of $N=(2,2)$ superconformal symmetries in the Hamiltonian formalism 
   and calculated their Poisson brackets (\ref{bliablia282})-(\ref{2202wkkwk2h2ss}).  These generators correspond to the Hamiltonian 
    realization of the superconformal symmetries of the  action
      (\ref{classical-action}).  One trivial, but important, 
     aspect is that coordinate changes $\tilde{x}^\mu= g^\mu(x)$ (with inverse $x^\mu =f^\mu(\tilde{x})$) can be mapped 
      to symplectomorphisms of (\ref{38383899}):
    \beq
      \tilde{\phi}^\mu = g^\mu (\phi)~,~~~~~~~~~~~~\tilde{S}_\mu=  \frac{\d f^\mu}{\d \tilde{\phi}^\nu} (g (\phi)) S_\nu~.
    \eeq{claskskew39399}
     Thus it is very easy to construct local expressions which are diffeomorphism invariant and calculate their Poisson 
      brackets. 
    
 In section \ref{CY} we considered the chiral de Rham complex
 $\Omega_{\hbar}^{ch}(M)$ on a Calabi-Yau manifold $M$. $\Omega_{\hbar}^{ch}(M)$
  is a sheaf of SUSY vertex algebras which are defined locally by the
  commutator: 
\beq
 [\phi^\mu (\xi, \theta),  S_\nu(\xi', \theta') ] = \hbar \delta^\mu_\nu ~ \delta (\xi- \xi') \delta (\theta-\theta')~,
 \eeq{xxjxkee9e9e00}
  which can be understood as a canonical quantization\footnote{Here we miss
  the factor ``$i$'' which is important for unitarity issues. 
     In the present consideration we proceed formally and ignore the hermiticity of the operators.} of the the structure (\ref{38383899}). 
  As in any operatorial quantization  we have to deal with the  ordering problem.   In CDR, we choose the standard normal 
    ordering for all operators.  Once we choose the ordering, we can look at the normal ordered version of (\ref{claskskew39399})
     (see (\ref{quantumtragshjj})) and ask if it respects the commutator
     (\ref{xxjxkee9e9e00}). As it stands, this is a complicated issue and it requires 
 additional care. CDR gives meaning and a solution to this question.  
      Thus, CDR provides a prescription by which to perform a canonical quantization of the non-linear sigma model. 
    Next, we look at the normally ordered expressions for the classical $({\cal P}, {\cal J}_1, {\cal J}_2, {\cal H})$ and it turns out that 
     ${\cal P}$ and ${\cal J}_1$ are not well-defined under ``quantum
     diffeomorphisms'' unless they are corrected by $\hbar$-terms. 
      In order to be able to  do this we require that  $M$ is a
      Calabi-Yau manifold.  Once we have well-defined operators, we can calculate 
       the operator algebra they generate, which turns out to be a
       central extension of the $N=(2,2)$ superconformal algebra. 

Therefore, CDR suggests to perform a canonical quantization locally on
$M$ and then glue each patch in an appropriate sense. 
 Indeed, if $M$ is an affine space with its standard flat metric, then the
 canonical  quantization of the sigma model can be carried out completely.
    The reader may consult the next section for explicit formulas. In
    fact, the free $\beta\gamma b c$-system (with an appropriate choice
    of Hamiltonian) can be interpreted
     as the canonical quantization of the sigma model with flat metric
     over an affine space.

            The set of global sections $V_\hbar := \Gamma(M,
	    \Omega_{\hbar}^{ch}(M))$ is a family of vertex algebras
	    associated to the manifold $M$ (depending on the parameter
	    $\hbar$). 
       The operators from $V_\hbar$ are interpreted as operators of the quantum sigma model in
        the Schr\"odinger picture (i.e., the operators are taken at fixed
	time, say $t=0$).  If we choose a Hamiltonian
          $H \in V_\hbar$, and postulate the flow equation (\ref{floweqs}),
            then we can work in the Heisenberg picture
	    (\ref{flowsolution}).  From this point of view, CDR does not carry any dynamical 
             information unless a Hamiltonian is introduced. This is
	     consistent for example with results from
	     \cite{Frenkel:2008vz}, where the authors derive CDR as the phase space
	for a suitable limit of the sigma-model.
 
  The structures which we find in CDR can be equivalently found with
  the path integral approach. CDR carries information  
 about the equal time commutators, while the path integral calculates the
 time ordered correlators:
\beq
 \langle T( O_1(\xi, t), O_2 (\xi', t')) \rangle = \int D\phi DS ~O_1(\xi, t) O_2 (\xi', t') e^{\frac{1}{\hbar}\left ( \int dt d\xi   d\theta \, S_\mu \d_t \phi^\nu -   \int dt H \right )}~,
\eeq{dkdkdk222}
 where $T$ stands for the time ordering. 
 In order to calculate equal time commutators we have to use the
 Bjorken-Johnson-Low prescription\footnote{It is important that we do not
 have a boundary in this field theory. The canonical relations may be modified by boundary contributions, e.g. see \cite{Baulieu:2001fi}.}
\beq
 \langle [O_1(\xi, t), O_2 (\xi', t)] \rangle = \left (\lim\limits_{t\rightarrow t'+0} - \lim\limits_{t\rightarrow t'- 0} \right ) \langle T( O_1(\xi, t), O_2 (\xi', t')) \rangle~.
\eeq{qualtimecommutator}
  The result of this calculation is independent from the Hamiltonian $H$. We can use this point splitting procedure to 
   define our local operators and study their properties under
   diffeomorphisms of the target manifold. 

   Let us a make a few concluding remarks. Our manipulations are formal and it is open to further investigation 
    how the Hamiltonian interpretation of CDR can  help us to understand the quantum sigma model.
      Especially one may need extra input for  the analysis of the solutions of the flow equation,  for example in order to make sense of
       some analytical issues. 
    Eventually, we would like to understand how to calculate and study the properties of 
     the non-equal time correlators in the full quantum theory. 
    
In the present considerations $M$ is assumed to be a simply connected
manifold. If $M$ is not simply connected then 
 the phase space $T^*{\cal L}M$ is not connected and thus we have to face
 further complications in its quantization. Moreover, even if $M$ is
 simply connected, $\mathcal{L}M$ may fail to be simply connected. In
 this situation, one can replace $M$ by a $U(1)$ gerbe with connection on
 $M$, or more generally by any Courant algebroid on $M$. CDR is replaced
 in this setup by the sheaves of SUSY vertex algebras studied in
 \cite{Heluani2} and \cite{Heluani:2008hw}. 

\subsection{Mathematical aspects}
\label{math}
In this section we briefly review the \emph{quasiclassical limit} of the
chiral de Rham complex. For a thorough introduction to Poisson vertex
algebras as limits of vertex algebras and the corresponding Hamiltonian
equations we refer the reader to \cite{kacdesole} (see also \cite[$\S
16.2$]{frenkelzvi}). For an extensive study of sheaves of Poisson vertex
algebras and their relation to chiral de Rham complex see
\cite{Malikov2006}.

Suppose we are given a family of vertex algebra structures $V_\hbar$ on
the same vector space $V$. That is, we have a collection of operations
$A_{(n, \hbar)}B \in V$ for all $A,B \in V$. We may think of the family
$V_{\hbar}$ as a vertex algebra over the ring of power series
$\mathbb{C}[ [\hbar]]$ instead of simply $\mathbb{C}$. 
Suppose moreover that in the limit $V_0=\lim_{\hbar
\rightarrow 0} V_\hbar:= V_{\hbar}/\hbar V_{\hbar}$, the vertex algebra becomes
commutative, that is, all the fields 
\[Y_\hbar(A,\xi) = \sum_{n \in \mathbb{Z}} \xi^{-1-n} A_{(n,\hbar)}~,
\]commute modulo $\hbar$-terms. Equivalently, it follows from
(\ref{commutatorssw93}) that all the products $A_{(j,\hbar)}B$, $j \geq
0$, vanish
modulo $\hbar$-terms, and moreover, the normally ordered product is
commutative modulo $\hbar$-terms. 

Note that this in particular says that
the $\lambda$-bracket vanishes modulo $\hbar$, hence we can rescale our
commutators on $V_0$ to define a new operation:
\begin{equation}
	\{A_\lambda B\} = \lim_{\hbar \rightarrow 0}
	\frac{1}{\hbar} [A_{\lambda} B]_{\hbar}~,
	\label{eq:poisson2}
\end{equation}
or, equivalently
\begin{equation}
	\{A(\xi), B(\xi')\} = \lim_{\hbar \rightarrow 0}
	\frac{1}{\hbar} \sum_{j \geq 0} \frac{1}{j!}
	\bigl(A_{(j,\hbar)}B\bigr)
	(\xi') \partial_{\xi'}^j \delta(\xi -\xi')~.
	\label{eq:poisson3}
\end{equation}

$V_0$ with its commutative product 
\[A \cdot B := \lim_{\hbar \rightarrow 0}
A_{(-1,\hbar)}B~,\] its \emph{Poisson $\lambda$-bracket}
(\ref{eq:poisson2}), and its derivation
\[ \partial A := \lim_{\hbar \rightarrow 0} A_{(-2,\hbar)} |0\rangle~, \] 
acquires an extra
structure known as a \emph{Poisson Vertex algebra} and it is called the
\emph{quasiclassical limit} of the family $V_\hbar$.

Performing computations on $V_0$ is generally much simpler than
performing computations on $V_\hbar$. For example, the bracket
$\{_\lambda\}$
and the product $\cdot$ satisfy a simple Leibniz rule on $V_0$:
\begin{equation}
	\{A_\lambda B\cdot C\} = \{A_\lambda B\}\cdot C + B \cdot \{A_\lambda C\}~,
	\label{eq:poisson4}
\end{equation}
while the corresponding relation on $V_\hbar$ -- known as the
non-commutative Wick formula, reads (we omit the $\hbar$ in the
operations):
\begin{equation}
	[A_\lambda :B C:] = :[A_{\lambda}B] C: + :B[A_\lambda C]: +
	\int_0^\lambda [[A_\lambda B]_\gamma C] d\gamma~.
	\label{eq:poisson5}
\end{equation}
Note that the integral term, under our hypothesis, vanishes of order
$\hbar^2$ since it is a double commutator. In particular, this
\emph{quantum correction} disappears in the limit $\hbar \rightarrow 0$
even after rescaling the OPE as in (\ref{eq:poisson2}). It is in this sense 
that one usually refers to Poisson vertex algebras as ``vertex algebras
with the quantum corrections removed''.

In section \ref{review} we constructed a family of sheaves of vertex
algebras $\Omega^{\mathrm{ch}}_\hbar(M)$ on a manifold $M$. Taking global
sections one obtains a family of vertex algebras $V_\hbar := \Gamma(M,
\Omega^{\mathrm{ch}}_\hbar(M))$. And finally, we obtain a Poisson vertex
algebra by the above limiting procedure: $V_0 := \lim_{\hbar \rightarrow
0} V_\hbar$. In fact, one can first take the limit locally, to obtain a
sheaf $\mathcal{V}_0$ of Poisson vertex algebras, and then taking global
sections we find $V_0 \simeq \Gamma(M, \mathcal{V}_0)$. Indeed, in a local
coordinate chart $\{x_\mu\}$, 
$\mathcal{V}_0$ is generated by fields (cf. (\ref{betagannaregs}) and
(\ref{betahaahhsbc}))  $(\beta_\mu, \gamma^{\mu})$ with their Poisson
bracket:
\begin{equation}
	\{\beta_{\mu}(\xi), \gamma^\nu(\xi')\} = \delta^\nu_\mu \delta(\xi -
	\xi')~, 
	\label{eq:poisson6}
\end{equation}
or equivalently
\begin{equation}
	\{{\beta_\mu}_\lambda \gamma^\nu\} = \delta^\nu_\mu~, 
	\label{eq:poisson7}
\end{equation}
together with the fermionic fields $(b_\mu, c^\mu) $ satisfying the
Poisson bracket\footnote{this is a $\mathbb{Z}_2$ graded
bracket.}:
\begin{equation}
	\{b_\mu(\xi), c^\nu(\xi')\}= \delta^\nu_\mu \delta(\xi-\xi')~,
	\qquad \text{or}\quad \{ {b_\mu}_\lambda c^\nu\} = \delta_\mu^\nu~.
	\label{eq:poisson8}
\end{equation}
One obtains other fields of $\mathcal{V}_0$ using the commutative product
$\cdot$ and the derivation $\partial$.

The above said can be easily generalized to the case when the fields
depend on an extra odd coordinate $\theta$, namely, when we have a family
$V_\hbar$
of SUSY vertex algebras, such that it becomes commutative in the limit
$V_0 = \lim_{\hbar \rightarrow 0} V_\hbar$. Again we find that
$\mathcal{V}_0$ is
generated by even superfields $\phi^\mu(\xi, \theta) = \gamma^\mu(\xi) +
\theta c^\mu(\xi)$ and odd superfields $S_\mu(\xi, \theta) = b_\mu(\xi) +
\theta \beta_\mu(\xi)$ satisfying the super-Poisson bracket:
\begin{equation}
	\{ {\phi^\mu}(\xi,\theta), S_\mu(\xi', \theta')\} =
	\delta_\mu^\nu \delta(\xi-\xi') \delta(\theta-\theta')~.
	\label{eq:poisson9}
\end{equation}
We see that this coincides with (\ref{susyddhhdpp}), thus it is not
surprising that formulas computed at the quantum level in CDR, when
viewed modulo $\hbar$-terms, that is, at the quasiclassical level,
coincide with the formulas of section \ref{sigma}.

Under a change of coordinates
$\tilde{x}^\mu = g^\mu(x)$ with its inverse $x^\mu=f^\mu(\tilde{x})$, the
superfields $\phi^\mu$ and $S_\mu$ transform as in
(\ref{quantumtragshjj}), but the multiplication is now commutative and
associative. In particular, $\mathcal{V}_0$ is a (infinite rank) vector bundle on
$M$. Note however, that even though the generating sections of
$\mathcal{V}_0$
transform in a tensorial manner, not all local sections do, as the
example of $DS_\mu = (\partial_\theta + \theta\partial_\xi)
S_\mu(\xi,\theta)$ shows. 

Using the above mentioned formalism, we can now interpret the results of
section \ref{sigma} as a classical limit of the equations of section
\ref{CY}. In fact, the expression for $\mathcal{P}$ in (\ref{FFFFF1})
defines a field of $V_0$, which is now interpreted as the limit of the
fields of 
$V_\hbar$ defined by (\ref{slslH}). Similarly, one obtains  all the
generators of $N=(2,2)$ superconformal symmetries of the classical sigma
model on a Calabi-Yau manifold $M$
(\ref{FFFFF1})-(\ref{FFFFF4}), as the limit of the corresponding sections
of CDR on $M$.

Given a Poisson Vertex algebra $V_0$ one usually considers
elements of $V_0/\partial V_0$ as
``local functionals'' \cite[Rem 6.3]{kacdesole}. And for a given local
functional $H$, one considers the ``Hamiltonian equations'' 
\begin{equation}
	\dot{u} = \{H, u\}~,~~~~~~~~~~ u \in V_0~.
	\label{eq:hamil1}
\end{equation}
In a similar manner, for a given vertex algebra $V$ and an element $H \in
V/\partial V$ we can construct the ``quantum Hamiltonian equations'' 
\begin{equation} \dot{u} =
	\frac{1}{\hbar} [H, u]~,~~~~~~~~~~ u \in V~. \label{eq:hamil3} \end{equation} 
Let $M$ be a Calabi-Yau manifold and let $V_\hbar$ be the vertex algebra of
global sections of CDR, $V_0$ its quasiclassical limit described
above. Recall that we have the global section $\mathcal{H}$ (\ref{H'}),
and this clearly gives rise to an element of $V_0$ as well. We consider the Hamiltonian 
\begin{equation}
	H= \frac{1}{2} \int d\sigma dt~  \mathcal{H}~.
	\label{eq:hamil2}
\end{equation}
For this choice of Hamiltonian in $V_0$, the equations (\ref{eq:hamil1})
agree with the equations of motion (\ref{ejdjw02020273}), derived in the
context of the classical non-linear sigma model. Therefore we see that the classical limit of the
chiral de Rham complex of $M$ together with the choice of
Hamiltonian (\ref{eq:hamil2}) recovers the full dynamics of the classical
non-linear sigma model on $M$.

Note however that since $\mathcal{H}$ is a well defined section in
$V_\hbar$ for all $\hbar$, we can use $H$ as a quantum Hamiltonian to
write down some quantum equations of motion. We refrain from doing so
here since these equations are not enlightening.   

\begin{remark}
	We have remarked earlier that the section (\ref{H'}) was constructed
	in \cite{Heluani2} under the assumption that $M$ is Calabi-Yau.
	Even though we conjecture that this section is well defined on
	any Riemannian manifold $M$, for the purposes of this section, we
	only need its zero mode $H$. It is straightforward to check that
	$H$ is well defined regardless of whether $\mathcal{H}$ is.
\end{remark}
\begin{remark}
	Note that by solving formally (\ref{eq:hamil3}) for each $u \in
	V$ and by using the state-field correspondence $Y$ of $V$, we can
	associate a ``field of two variables''\footnote{Here we obviate
	the odd coordinates for simplicity.}
	\begin{equation}
		u \mapsto Y(u,\sigma,t) := Y(u(t), \sigma).
		\label{eq:2coords1}
	\end{equation}
	Following \cite{frenkelzvi} we can formally use the Virasoro
	$L_L$ (resp. $L_R$) 
	from section \ref{CY} to see how these ``fields'' change
	with respect to changes in the coordinate $\sigma + t$ (resp.
	$\sigma - t$), but a priory we do not know how to deal with
	general coordinates in the worldsheet. The appropriate algebraic
	axioms satisfied by the fields (\ref{eq:2coords1}) will be
	studied elsewhere.
\end{remark}

\section{Free field examples}
\label{free}

In this section we consider  the free field examples and for simplicity we set $\hbar=1$.
 We remind some standard facts about $\beta\gamma$-  and $bc$-systems and also stress some  aspects 
  of these systems in light of our previous discussion. 
 Although we here repeat explicitly some of the formulas from section \ref{CY}  we find this discussion  instructive 
  and clarifying. 

\subsection{Free boson}
\label{boson}

Consider a single  $\beta\gamma$-system defined by the bracket
\beq
 {[\beta}_\lambda \gamma] = 1~,
\eeq{betaggamma}
 where we use the $\lambda$-bracket notation, see section \ref{review}. 
 The Virasoro field is defined as:
\begin{equation} \label{bglbracket}
L=\beta \partial \gamma~,
\end{equation}
  which satisfies the Virasoro algebra:
\begin{equation}
{[L}_\lambda L]=(2\lambda+\partial)L+\frac{2\lambda^3}{12},
\end{equation}
 with central charge 2. With respect to this $L$, $\beta$ is a primary field of conformal weight $1$ and $\gamma$ is 
  of conformal weight $0$,
  \beq
   \gamma(\xi) = \sum\limits_{n \in \mathbb{Z}} \gamma_n \xi^{-n}~,~~~~~~~~~~~
   \beta(\xi) = \sum\limits_{n \in \mathbb{Z}} \beta_n \xi^{-n-1}~. 
  \eeq{expansionss122}
     The vacuum is annihilated by $\beta_n$ $(n \geq 0)$ and $\gamma_n$ $(n >0)$, but $\gamma_0$ is considered 
     to be a creator.     The field $L$ can be split into two parts as follows:
\begin{equation}
 L_{L} =\frac{1}{2}\left[\beta \partial \gamma + \frac{1}{2}\left(\beta^2+(\partial \gamma)^2\right)\right]~,~~~~
  L_{R} =\frac{1}{2}\left[\beta \partial \gamma - \frac{1}{2}\left(\beta^2+(\partial \gamma)^2\right)\right]~.
 \end{equation}
 $L_L$ and $L_R$  give rise to two commuting copies of the  Virasoro algebra with central charge $c=1$ each:
\begin{equation}
\begin{split}
{[L_{L}}_\lambda L_{L}]&=(2\lambda+\partial)L_{L}+\frac{\lambda^3}{12} ~,\\
{[L_{R}}_\lambda L_{R}]&=(2\lambda+\partial)L_{R}+\frac{\lambda^3}{12} ~,\\
{[L_{L}}_\lambda L_R]&=0~.
\end{split}
\end{equation}
 Neither $\beta$ nor $\gamma$ are primary fields with respect to $L_L$ and $L_R$.  However the field 
 $\tfrac{1}{\sqrt{2}}(\beta + \d \gamma)$ (resp. $\tfrac{1}{\sqrt{2}}
 (\beta - \partial \gamma)$) is primary of conformal weight $1$ with respect to 
 $L_{L}$ (resp. $L_R$).   Indeed, introducing new fields
 \begin{equation} \label{bgbmap}
\alpha_\pm=\frac{\beta \pm \partial\gamma}{\sqrt{2}}~, 
\end{equation}
  which satisfy the following brackets:
\begin{equation}
{[\alpha_\pm}_\lambda\alpha_\pm] = \pm \lambda~, \quad
{[\alpha_\pm}_\lambda \alpha_\mp]=0~,
\end{equation}
  the Virasoro fields $L_{L/R}$ can be written as:
\begin{equation}
L_L=\frac{1}{2}\alpha_+^2 ~,\quad
L_R=-\frac{1}{2} \alpha_-^2~.
\end{equation}
 These are just formal observations about the vertex algebra of the $\beta\gamma$-system. 
 
 Next, let us introduce a new parameter $t$, and a $t$-dependence on our
 fields by means of flow equations. 
  We choose the Hamiltonian to be:
\begin{equation}\label{ssj3993939www}
H= i \int d\xi \,\xi \left(L_L - L_R \right) = \frac{i}{2} \int
d\xi\,\xi\,\left(\beta^2+(\partial \gamma)^2\right)~,
\end{equation}
or equivalently, in the $\sigma$-coordinates $\xi=e^{i\sigma}$:
\begin{equation}
H = \frac{1}{2} \int d\sigma\,\left(\beta^2+(\partial_\sigma \gamma)^2\right)~.
\end{equation}
 In performing this coordinate change we have used that $\beta$ is of
 conformal weight one and $\gamma$ is of conformal weight zero, and we
 use the ``diagonal Virasoro'' $L= L_L + L_R$ to perform
 changes of coordinates. 
 The flow equations are:
 \beq
  \frac{d\gamma}{d t} = \beta~,~~~~~~~~~\frac{d \beta}{d t} = \d_\sigma^2
  \gamma~,
 \eeq{flowbetahhha022}
 and these are exactly the Hamiltonian equations for the free boson theory. The solution of (\ref{flowbetahhha022})
 is given by the following expression: 
  \beq
  \gamma(\sigma, t) = \gamma_0 + \beta_0 t + \sum\limits_{n\neq 0} \frac{1}{2}\left [ (\gamma_n +
    \frac{i}{n}\beta_n) e^{-in (\sigma +t)}
   + (\gamma_n - \frac{i}{n}\beta_n) e^{-in (\sigma -t)} \right ]~.
  \eeq{solutionbison}
  Thus, the $\beta\gamma$-system with an appropriate choice of Hamiltonian can be identified with 
   the standard free boson theory, which contains both chiral and anti-chiral sectors.
    Among the solutions to the  flow equation, the field $\alpha_+$
    corresponds to the left moving (chiral) sector, 
    while $\alpha_-$ to the
      right moving  (anti-chiral) sector.  It is crucial for the identification with the free boson that $\gamma_0$ is a creator
      so that we have states: 
 \beq
  e^{ik \gamma_0} |0 \rangle \equiv |k \rangle~,~~~~~~~~~~\beta_0
  |k\rangle = ik |k \rangle~,~~~~~~~~~~k\in \mathbb{R}~.
 \eeq{feu33momnetusm} 
 
Let us finish with a simple but important comment. If we choose a
different Hamiltonian, then we 
 obtain different flow equations and a different system. For example, if we consider the Hamiltonian
\beq
 H = \int d\sigma~\beta \d_\sigma \gamma~,
\eeq{hasmmms20202}
the flow equations are 
\begin{equation}
\frac{d \gamma}{d t}=\partial_\sigma \gamma~, \quad
\frac{d \beta}{d t}=\partial_\sigma \beta~,
\end{equation}
 which  is just the standard chiral $\beta\gamma$-system.

\subsection{Free fermion}
\label{fermion}
 Next, we consider the super vertex algebra of the free $bc$-system,
 which is generated by two odd fields $b$ and $c$ satisfying the
 following bracket:
\begin{equation} \label{bclbracket}
{[b}_\lambda c]=1~.
\end{equation}
 The field
\beq
L =\frac{1}{2}\Bigl( (\partial c)b+ (\partial b)c\Bigr)~,
\eeq{vsie3494949}
gives rise to a Virasoro algebra of central charge $1$. With respect to
$L$, $c$ and $b$ are primary 
 fields of conformal weight $1/2$:
 \beq
  c(\xi) = \sum\limits_{n \in \mathbb{Z}+1/2} c_n \xi^{-n-\frac{1}{2}}~,~~~~~~~~~~~
   b(\xi) = \sum\limits_{n \in \mathbb{Z}+1/2} b_n \xi^{-n-\frac{1}{2}}~. 
 \eeq{ckdkdkdkkdkdk2w22}
  The vacuum is annihilated by $b_n$ and $c_n$ $(n > 0)$ and the rest of
  the modes are creators.  $L$ can be split  into two commuting parts:
\begin{equation}
L_{L/R} = \frac{1}{4}\Bigl( (\partial c)b+ (\partial b)c \pm (\partial
c)c \pm (\partial b)b\Bigr) ~,
\end{equation}
such that each $L_{L/R}$ generate a Virasoro vertex algebra with central
charge $\frac{1}{2}$ and they mutually commute. 
  Introducing the new fields:
\begin{equation} \label{bcfmap}
\psi_+ =\frac{b+c}{\sqrt{2}}~, \quad
\psi_- =\frac{b-c}{\sqrt{2}}~, 
\end{equation}
 satisfying the brackets
\begin{equation}
{[\psi_\pm}_\lambda\psi_\pm]=\pm1 ~,\quad
{[\psi_\pm}_\lambda \psi_\mp]=0~,
\end{equation}
we rewrite $L_{L/R}$ as:
\begin{equation}
L_L=\frac{1}{2} (\partial\psi_+) \psi_+ ~,\quad
L_R=-\frac{1}{2} (\partial \psi_-) \psi_- ~.
\end{equation}
Introducing the Hamiltonian:
\beq
 H = i \int d\xi\, \xi\, \left (L_L - L_R \right )= \frac{i}{2} \int
 d\xi\, \xi\, \left ( (\partial c)c + (\partial b)b\right)  = \frac{1}{2}
 \int d\sigma  \left ( (\partial_\sigma c)c 
  + (\partial_\sigma b)b\right),
\eeq{newieie939933}
 the flow equations imply:
 \beq
  \frac{d \psi_\pm}{d t} \mp \d_\sigma \psi_\pm =0 ~, 
  \eeq{ddhfhfhhfhhfh333}
and we can recognize these as the left and right moving parts  of   the
standard free periodic fermion system on the cylinder. 

\subsection{N=1 supersymmetry}
 We can combine the  $\beta\gamma$- and $bc$-systems from the previous subsections. 
  The fields:
\begin{equation}
L=\beta \partial \gamma+\frac{1}{2}\left( (\partial c)b+ (\partial b)c\right)~,
\quad
G=c\beta+ (\partial \gamma) b,
\end{equation}
give rise to the $N=1$ superconformal algebra with central charge $3$:
\begin{equation}
\begin{split}
{[L}_\lambda L]&=(2\lambda+\partial)L+\frac{3\lambda^3}{12}~, \\
{[L}_\lambda G]&=(\partial + \frac{3}{2}\lambda)G ~,\\
{[G}_\lambda G]&=2L+ \lambda^2~.
\end{split}
\end{equation}
Introducing the superfields $\phi = \gamma + \theta c$ and $S= b+ \theta
\beta$ we have (cf. (\ref{slslH})):
\beq
 {\cal P} = G + 2 \theta L = D\phi DS + (\d \phi) S~,
\eeq{wwiwiw2002020}
 With respect to $L$,  $\phi$ is a primary field of conformal weight $0$ and $S$ is primary of 
  conformal weight $1/2$. 

 We can split $L$ and $G$ into two copies,
\begin{equation}
\begin{split}
 L_{L/R} & =\frac{1}{2}\left ( \beta \partial \gamma \pm \frac{1}{2}\left(\beta^2+(\partial \gamma)^2\right)
 + \frac{1}{2}\left( (\partial c)b+ (\partial b)c \pm (\partial c)c \pm
 (\partial b)b\right) \right ), \\
G_{L/R}  & = \frac{1}{2} \Bigl( c\beta+ (\partial \gamma) b \pm \beta b
\pm (\partial \gamma) c \Bigr), 
\end{split}
\end{equation}
of commuting $N=1$ superconformal algebras with central charge $3/2$ each. Using superfields 
 we can combine 
 \beq
  G_{L/R} + 2\theta L_{L/R} = \frac{1}{2} \left ( D\phi DS + \d \phi S
  \pm \d \phi D\phi \pm (DS) S \right )~. 
 \eeq{48484jejejeje}
 It is important to stress that the fields $\phi$ and $S$ are not primary
 with respect to $L_{L/R}$, but rather their sum $L= L_L +L_R$.
  The Hamiltonian is just the sum of the expressions (\ref{ssj3993939www}) and (\ref{newieie939933}) and the corresponding 
 flow equations realize the free boson-fermion system. 
 
We can consider $d$ copies of the $\beta\gamma bc$-system.
  The corresponding $N=1$ superconformal algebra of central charge $3d$
  is generated by:
\ber
\label{gopa11}L&=& (\partial\gamma^\mu) \beta_\mu +\frac{1}{2}\Bigl(
(\partial b_\mu) c^\mu + (\partial c^\mu) b_\mu \Bigr)~, \\
\label{gopa22}G&= & (\partial\gamma^\mu) b_\mu + (c^\mu)\beta_\mu ~.
\eer{susyN1Rn}
  For a constant $d\times d$ symmetric non-degenerate matrix  $g_{\mu\nu}$ we define 
   the fields
\ber
\label{supergopa1}\alpha_+^\mu = \frac{g^{\mu\nu}\beta_\nu + \d \gamma^\mu}{\sqrt{2}}~,&~~~~~~~~~&
\alpha_-^\mu = \frac{g^{\mu\nu}\beta_\nu - \d \gamma^\mu}{\sqrt{2}}~,\\
\label{supergopa2}\psi_+^\mu = \frac{g^{\mu\nu}b_\nu + c^\mu}{\sqrt{2}}~,&~~~~~~~~~&
\psi_-^\mu = \frac{g^{\mu\nu}b_\nu - c^\mu}{\sqrt{2}}~,
\eer{febosreedidnwoow}
 with  brackets
 \ber
 {[\alpha^\mu_\pm}_\lambda \alpha^\nu_\pm] = \pm g^{\mu\nu} \lambda~,&~~~~~~~~~~~~~~~&  {[\alpha^\mu_+}_\lambda \alpha^\nu_-] = 0~,\\
 {[\psi^\mu_\pm}_\lambda \psi^\nu_\pm] = \pm g^{\mu\nu} ~,&~~~~~~~~~~~~~~~&  {[\psi^\mu_+}_\lambda \psi^\nu_-] = 0~.
\eer{basi20200}
 The $N=1$ algebra (\ref{susyN1Rn}) can be split into two commuting
 copies $L= L_L + L_R$ and $G= G_L + G_R$ \cite[Ex. 5.9a]{Kac:1996wd},
\ber
\label{LLRdkdkd} L_{L/R} &=& \pm \frac{1}{2} \alpha^\mu_\pm g_{\mu\nu} \alpha^\nu_\pm \pm \frac{1}{2} \d \psi_\pm^\mu g_{\mu\nu} \psi_\pm^\nu~,\\
G_{L/R} &=& \pm   \alpha^\mu_\pm g_{\mu\nu}  \psi^\nu_\pm~.
\eer{newsss12228282}
 Each has central charge $\frac{3}{2}d$. If we choose as Hamiltonian
 $i (L_L)_0 - i (L_R)_0$, we 
 obtain the quantization of the sigma model on $\mathbb{R}^d$ with metric $g_{\mu\nu}$. 
  
  On a general manifold $M$ we can glue together $\beta\gamma
  bc$-systems according to \cite{Malikov:1998dw}. When $M$ is
  orientiable,
   the generators (\ref{gopa11}) and (\ref{gopa22}) can be glued in a
   consistent manner (they are the components of 
   (\ref{slslH})).

\subsection{N=2 supersymmetry}
\label{freesusy} 
Next we consider $\mathbb{C}^n = \mathbb{R}^{2n}$ and 
we follow the notations from section \ref{CY}. If we choose a constant symplectic  form $\omega$ then
the fields $(J_2, G^\pm_2, L_2)$, with $L_2=L$ in (\ref{gopa11}),  $G_ 2^+ + G_2^- =G$ in (\ref{gopa22}) and
\ber
J_2&=&\frac{i}{2}\left(\omega^{\mu\nu}b_\mu b_\nu -\omega_{\mu\nu}c^\mu
c^\nu \right), \\
G_2^- - G_2^+& = & i\left(\omega^{\mu\nu}\beta_\mu
b_\nu-\omega_{\mu\nu}\partial\gamma^\mu c^\nu\right),
\eer{alslslsl30303}
 generate the $N=2$ superconformal algebra with central charge $6n$.  If
 we choose the standard constant complex
  structure $I$, then the fields $(J_1, G_1^\pm, L_1)$ with $L_1=L$ in (\ref{gopa11}),  $G_1^+ + G_1^- = G$
   in (\ref{gopa22}) and
 \ber
J_1&= &iI^{\mu}_{\nu}c^{\nu}b_{\mu},\\
G_1^- - G_1^+& = &
iI^{\mu}_{\nu}\partial\gamma^{\nu}b_{\mu}-iI^{\mu}_{\nu}c^{\nu}\beta_{\mu},
\eer{ksksksk3333}
generate the $N=2$ superconformal algebra with central charge $6n$.  If we require that $\omega$ and $I$
 are compatible, i.e. we consider the Hermitian metric $g= -\omega I$, then using (\ref{supergopa1})
  (\ref{supergopa2}) we can introduce two commuting copies of the $N=2$ algebra $(J_{L/R}, G^\pm_{L/R}, L_{L/R})$
   with central charge 
  $\frac{3}{2}d$ each.   Together with $L_{L/R}$  in (\ref{LLRdkdkd})
  the other fields are written as \cite[Ex. 5.9d]{Kac:1996wd} \beq
 G_{L/R}^+ = \pm \alpha_{\pm}^i g_{i\bar{j}} \psi_{\pm}^{\bar{j}}~,~~~~~~~~~
  G_{L/R}^- =  \pm \alpha_{\pm}^{\bar{i}} g_{\bar{i}j} \psi_{\pm}^{j}~,~~~~~~~~~
  J_{L/R}  = \pm  \psi^{\bar{i}}_\pm g_{\bar{i}j} \psi_\pm^{j},
\eeq{kxkd9993030}
 where we use holomorphic coordinates $(i, \bar{i})$. 

Here one can explicitly see that this is a canonical quantization of
the $N=(2,2)$ supersymmetric sigma model with target $\mathbb{C}^n$.  
 All operators above can be understood in the Schr\"odinger picture. Introducing the Hamiltonian 
  $i (L_L)_0 - i (L_R)_0$ we get the correct time dependence. 

\subsection{Topological A-and B-twist}\label{twist}
Here we briefly discuss the topological twist of the sigma model which we just discussed. All our formulas 
 are written in the flat $\mathbb{C}^n$ case, but they have a
 straightforward generalization for a  Calabi-Yau manifold. 

Starting with the left and right moving $N=2$ superconformal algebra  $(J_{L/R}, G^\pm_{L/R}, L_{L/R})$, 
 we can perform a  topological twist.  The idea is to redefine the Virasoro field in such a way that the central charge becomes zero. Given the algebras $(J_{L/R}, G^\pm_{L/R}, L_{L/R})$ we can define two inequivalent twists, 
 the A-twist and the B-twist. The A-twist is given by
\begin{equation}
\begin{split}
L_L & \rightarrow L_L+\frac{1}{2}\partial J_L~, \\
L_R & \rightarrow L_R-\frac{1}{2}\partial J_R~,
\end{split}
\end{equation}
and the B-twist by:
\begin{equation}
\begin{split}
L_L & \rightarrow L_L+\frac{1}{2}\partial J_L ~,\\
L_R & \rightarrow L_R+\frac{1}{2}\partial J_R~.
\end{split}
\end{equation}
 Using the notations from section \ref{CY}
   we have
   $J_{L/R}=\frac{i}{2}\left(\mathcal{J}_1\pm\mathcal{J}_2\right)|_{\theta=0}$,
   and recall that both $\mathcal{J}_1$ and $\mathcal{J}_2$ give rise to
   the same
   $L=L_L+L_R$. The A-twist corresponds to a twist with respect to $J_2=i\mathcal{J}_2|_{\theta=0}$, while the B-twist corresponds to a twist with respect to $J_1=i\mathcal{J}_1|_{\theta=0}$.

 Let us perform the A-twist. Consider the $N=2$ algebra $(J_2, G^\pm_2, L_2)$ 
  and redefine $L_2 \rightarrow L_2 + \frac{1}{2}\d J_2$. Defining the
  new fields:  
\ber
\psi_\mu &=&\frac{1}{\sqrt{2}}\left(b_\mu-i\omega_{\mu\nu}c^\nu\right), \\
\chi^\mu & =&\frac{1}{\sqrt{2}}\left(c^\mu-i\omega^{\mu\nu}b_\nu\right), \\
 \beta_\mu'&=&\frac{\beta_{\mu}-i\omega_{\mu\nu}\partial\gamma^{\nu}}{\sqrt{2}}
 ,\\
\gamma'^{\mu}&=&\sqrt{2}\gamma^{\mu},
\eer{ssksksww99qq}
with the brackets
\beq
{[\beta'_\mu}_\lambda \gamma'^\nu] = \delta^\nu_\mu~,~~~~~~~~~~~~~~~~
{[\chi^\mu}_\lambda \psi_\nu] = \delta^\mu_\nu
\eeq{djdjjd399930ssks}
we obtain a topological $N=2$ algebra $(L_2, J_2, Q_2, G_2)$: 
\ber
\label{supergopa11}L_2&=&\partial \gamma^\mu \beta_\mu + \partial\chi^\mu
\psi_\mu,\\
\label{supergopa22}J_2&=&\chi^\mu\psi_\mu, \\
\label{supergopa33}Q_2 &= & G_2^+ = \chi^\mu\beta_\mu, \\ 
\label{supergopa44}G_2 & =&  G_2^-=
\psi_\mu\partial\gamma^{\mu}-i\omega^{\mu\nu}\psi_\mu\beta_{\nu},
 \eer{ddjdkkeekeek}
 where we have omitted $'$ on $\beta$ and $\gamma$. 
With respect to this new Virasoro, the fields $\psi_\mu$ and  $\chi^\mu$
have conformal weights $1$ and $0$ respectively. Moreover, the odd generators $Q_2$ and $G_2$ have conformal weights $1$ and $2$, thus the zero mode of $Q_2$
 is a  BRST operator.
   Note that the last term in (\ref{supergopa44}) can actually be rewritten as 
\begin{equation}
\begin{split}
G_2 & = \psi_\mu\partial\gamma^{\mu}+[(Q_2)_0,V] ~,\\
V&=\frac{i\omega^{\mu\nu} \psi_\mu\psi_\nu}{2}~.
\end{split}
\end{equation}
The expressions (\ref{supergopa11})-(\ref{supergopa44}) 
 look like those in \cite{Malikov:1998dw} modulo this BRST exact
term.  This whole construction can be carried out on any symplectic
manifold, where we can use Darboux coordinates. By the usual
arguments, the BRST cohomology 
  is concentrated in conformal weight zero, and the class of 
 \begin{equation}
\mathcal{O}_A=A_{\mu_1...\mu_k}(\gamma)\chi^{\mu_1}...\chi^{\mu_k}~,
\end{equation}
 is identified with the class $[A] \in H^k_{dR}(M)$, the de Rham
 cohomology of $M$.

The B-twist corresponds to twisting the $N=2$ algebra $(J_1, G^\pm_1, L_1)$.
  Redefining $L_1 \rightarrow L_1 + \frac{1}{2}\d J_1$ and using
  holomorphic coordinates,
   the corresponding topological $N=2$ algebra $(L_1, J_1, Q_1, G_1) $ is given by
 \ber
L_1&=&\partial\gamma^i \beta_i + \partial\gamma^{\bar{i}} \beta_{\bar{i}} +\partial b_{i} c^{i} + \partial c^{\bar{i}} b_{\bar{i}}~, \\
J_1&=&b_{i}c^{i}-b_{\bar{i}}c^{\bar{i}}~, \\
Q_1 &=& G_1^+=\partial\gamma^{i}b_{i}+c^{\bar{i}}\beta_{\bar{i}}~,\\ 
G_1 & =& G_1^- =\partial\gamma^{\bar{i}}b_{\bar{i}}+c^{i}\beta_{i}~.
\eer{bliasuki}
With respect to the new Virasoro, the fields $b_{i},
b_{\bar{i}},c^{i},c^{\bar{i}}$ have  conformal weights $0,1,1,0$
respectively. The two odd generators $Q_1$ and $G_1$ are of conformal weight $1$ and $2$. Thus the zero mode 
 of   $Q_1$ is  a BRST operator. Using holomorphic coordinates we can
 glue all these formulas on a
  Calabi-Yau manifold. The BRST cohomology  is concentrated in conformal
  weight zero, and the class of  
\begin{equation}\mathcal{O}_B=B_{\bar{i}_1...\bar{i}_p}^{j_1...j_q}(\gamma)c^{\bar{i_1}}...c^{\bar{i}_p}b_{j_1}...b_{j_q}
\end{equation}
  can be identified with  the class $[B] \in H^{p}_{\bar{\d}} (M,
  \wedge^q T^{1,0}_M)$. 
  \section{Summary}
\label{end}

 In this work we suggested to interpret $\Omega_\hbar^{ch}(M)$ as the prescription for a canonical quantization 
  of the non-linear sigma model\footnote{Also see the related work by Malikov  \cite{Malikov2006} on the relation of 
  CDR and the Lagrangian approach.}.  In \cite{Heluani2} it has been shown
  that for a Calabi-Yau manifold $M$ with a fixed  Ricci 
   flat metric one can construct global sections of CDR which generate   two commuting copies of $N=2$ 
    superconformal algebra with central charge  $\frac{3}{2} \dim_{\mathbb{R}} M$ each.  The form of these global 
     sections has been guessed in \cite{Heluani2}. Here we show that these sections can be derived systematically from the 
      the sigma model action. Namely, they correspond to the Hamiltonian realization of $N=(2,2)$ superconformal 
       symmetry for the sigma model with a Calabi-Yau manifold as target.  We explain the matching of classical and quantum brackets, 
        thus supporting the idea of the Hamiltonian interpretation of CDR.  In the Hamiltonian formalism, the dynamics is 
         introduced through the flow equations upon the choice of a
	 Hamiltonian function. In fact, if we choose the correct
	 Hamiltonian, CDR contains the classical 
          dynamics  of the non-linear sigma model.  Amazingly this is the same Hamiltonian 
           which allows us to interpret the two copies of the $N=2$ algebra in CDR as left (chiral) and right (anti-chiral) sectors. 
           
     Let us make a side remark at this point. In the context of our interpretation, the ability to construct
       two commuting  copies of $N=2$ (with the correct central charges!)
       attached to a Calabi-Yau 
      manifold with a Ricci flat metric raises a puzzle about the multi-loop calculations for the supersymmetric sigma model.
       It is believed that the sigma model on a Calabi-Yau manifold  is
       superconformal, but for a non Ricci-flat metric.
         The reader may consult  \cite{Nemeschansky:1986yx} for details.  We hope to address this puzzle elsewhere. 
           
   The proposed interpretation of CDR also suggests some interesting
   mathematical ideas for vertex algebras. A   
   vertex algebra  can be understood as a quantum field theory in the
   Schr\"odinger picture. In particular, the 
  $\beta\gamma bc$-system can be interpreted as a free boson and free fermion in the Schr\"odinger picture. 
   Once the appropriate Hamiltonian is introduced,  we can switch to the Heisenberg picture by means of the flow
    equations. We hope that this idea may allow to incorporate both chiral and anti-chiral sectors in a single 
     mathematical framework. 
     
    CDR should describe many systems which have $T^*{\cal L}M$ as a classical phase space.
    The dynamics would crucially depend on the concrete choice of 
     Hamiltonian.  Moreover, we can deal with the gauge systems on $T^*{\cal L}M$ (e.g., the Poisson sigma model, the non-linear model 
      on $M/G$ when $M$ admits the action of a Lie group $G$, etc.). The appropriate  BRST symmetries on CDR should be introduced 
       to take care of the gauge symmetries,   very much in the spirit of  \cite{Borisov:1998dw}. 
     
     For example, in order to describe the system related to
     the large volume limit of the sigma model
      \cite{Frenkel:2005ku, Frenkel:2006fy, Frenkel:2008vz} we have to
      choose the following Hamiltonian:
      $$ H = i \int d\xi d\theta \xi ~(\d \phi^i S_i - \d \phi^{\bar{i}} S_{\bar{i}})~,$$
       written in holomorphic coordinates. One can easily check that the
       corresponding flow equations admit 
        holomorphic maps as solutions.   We  hope that one can derive the results from  \cite{Frenkel:2005ku, Frenkel:2006fy, 
        Frenkel:2008vz} in the present operatorial Hamiltonian framework.
	Indeed, it would be a good playground to check
        if we can describe the instanton correction within CDR with the
	Hamiltonian flow equations.

\bigskip\bigskip

\noindent{\bf\Large Acknowledgement}:
\bigskip

\noindent  We thank Edward Frenkel,  Andrei Losev, Joseph Minahan and
Konstantin Zarembo for illuminating discussions. 
  RH and MZ thank KITP, Santa Barbara where part of this work was 
carried out.  The research of RH and MZ was supported in part by DARPA under Grant No.
HR0011-09-1-0015 and by the National Science Foundation under Grant
No. PHY05-51164. The research of RH is supported by the Miller Institute for basic research
in Science.
The research of MZ is supported by VR-grant 621-2008-4273. 
\appendix
 
\Section{Quantum and classical brackets on Calabi-Yau}
\label{a:}

On a Calabi-Yau manifold $M$, equipped with a complex structure $I$ and a
symplectic structure $\omega$ such that $g=
-\omega I$ is a Ricci-flat metric, 
 we define  the  global sections of $\Omega_\hbar^{ch}(M)$ $({\cal P},
 {\cal J}_1, {\cal J}_2, {\cal H})$ 
 given by (\ref{slslH}), (\ref{3399skskkk}), (\ref{skskwow999}) and
 (\ref{H'}) respectively.   
  These sections form the following algebra with respect to the $\Lambda$-bracket \cite{Heluani2} :
\ber
\label{LLLL1} && {[{\cal P}}_\Lambda {\cal P}] = \hbar (2 \d+ \chi D + 3 \lambda){\cal P} + \hbar^2 \frac{c}{3}  \lambda^2 \chi~,\\
&&  {[{\cal P}}_\Lambda {\cal J}_1] = \hbar (2 \d + 2 \lambda + \chi D) {\cal J}_1~,\\
&&  {[{\cal P}}_\Lambda {\cal J}_2] = \hbar (2 \d + 2 \lambda + \chi D) {\cal J}_2~,\\
&&  {[{\cal J}_1}_\Lambda {\cal J}_1] = - \hbar {\cal P} - \hbar^2 \frac{c}{3} \lambda \chi~,\\
&&  {[{\cal J}_2}_\Lambda {\cal J}_2] = - \hbar {\cal P} - \hbar^2 \frac{c}{3} \lambda \chi~,\\
&&  {[{\cal J}_1}_\Lambda {\cal J}_2] = - \hbar {\cal H} ~,\\
&& {[{\cal H}}_\Lambda {\cal J}_1] =  \hbar (2 \d+ \chi D + 2 \lambda){\cal J}_2~,\\
 && {[{\cal H}}_\Lambda {\cal J}_2] =  \hbar (2 \d+ \chi D + 2 \lambda){\cal J}_1~,\\
  && {[{\cal H}}_\Lambda {\cal H}] =  \hbar (2 \d+ \chi D + 3 \lambda){\cal P} + \hbar^2 \frac{c}{3} \lambda^2 \chi~,\\
\label{LLLL10}  &&   {[{\cal P}}_\Lambda {\cal H}] =  \hbar (2 \d+ \chi D + 3 \lambda){\cal H} ~,
\eer{ddkd93030302--1}
 where $c= 3 \dim_{\mathbb R} M$. We introduce the following
 notations:
$Z=(\xi, \theta)$ and $Z'=(\xi', \theta')$, $\d'\equiv \d_{\xi'}$, $D'=
D_{\xi' \theta'}$,
  $\delta(Z, Z')= \delta(\xi-\xi') \delta(\theta-\theta')$.
  Then the above algebra for the operators $({\cal P}, {\cal J}_1, {\cal J}_2, {\cal H})$ can be written equivalently in 
   terms of commutators as 
\ber
\nonumber [{\cal P}(Z),  {\cal P}(Z')] & = &\hbar \left ( 2   \delta(Z,Z') \d'  -  D'\delta(Z,Z') D' + 3  \d' \delta(Z,Z') \right ) {\cal P}(Z')  \\
  \label{askssl111} &&-  \hbar^2 \frac{c}{3}  \d'^2  D' \delta(Z,Z')  ~,   \\
\label{1234abc}  [ {\cal P} (Z),  {\cal J}_i (Z')] & =& \hbar \left ( 2\delta(Z,Z') \d' + 2 \d' \delta(Z,Z')   -  D'\delta(Z,Z') D' \right ) {\cal J}_i(Z')~,\\
 \label{sjdkdkdkdaa} [{\cal J}_i(Z),  {\cal J}_i(Z')] &=& - \hbar \delta(Z, Z') {\cal P}(Z') + \hbar^2 \frac{c}{3} \d'  D'\delta(Z,Z')  ~,\\
 \label{33902001ss} [{\cal J}_1 (Z),  {\cal J}_2(Z')] & =& - \hbar \delta(Z,Z') {\cal H}(Z') ~,\\
\label{wui9999}  [{\cal H}(Z),  {\cal J}_1(Z')] &= &  \hbar (2 \delta(Z,Z')\d'- D'\delta(Z, Z') D' + 2 \d' \delta(Z, Z')){\cal J}_2(Z')~,\\
\label{sksks9292} [{\cal H}(Z), {\cal J}_2(Z')] & = &  \hbar (2 \delta(Z,Z')\d'- D'\delta(Z,Z') D' + 2 \d' \delta(Z,Z')){\cal J}_1(Z')~,\\
\nonumber  [{\cal H}(Z),  {\cal H}(Z')] & = &  \hbar (2\delta(Z,Z') \d' -D'\delta(Z,Z') D' + 3 \d' \delta(Z,Z')){\cal P}(Z') \\
  \label{22929wiiiw}  && - \hbar^2 \frac{c}{3} \d'^2 D' \delta(Z,Z') ~,\\
 \label{2202wkkwk2h2}   [{\cal P}(Z),  {\cal H}(Z')]  & =  &  \hbar (2\delta(Z,Z') \d'- D'\delta(Z, Z') D' + 3 \d' \delta(Z,Z')){\cal H} (Z') ~.
 \eer{didixkkxshhkaa}
  If we take the classical limit $\hbar \rightarrow 0$ and make the
  replacement $[~,~] \rightarrow \hbar \{~, ~\}$, then we can view this
  algebra as a super Poisson subalgebra of the 
  Poisson algebra of  functionals  on $T^* {\cal L}M$, the cotangent bundle of the superloop space.  Using the following 
  conventions for the local functional $F$:
 \beq
    \delta F = \int d\xi d\theta~  \left ( \frac{ F \overleftarrow{\delta}}{\delta S_\mu} \delta S^\mu +
  \frac{F \overleftarrow{\delta}}{\delta \Phi^\mu} \delta \Phi^\mu \right ) = \int d\xi d\theta
    \left ( \delta S_\mu \frac{ \overrightarrow{\delta} F}{\delta S_\mu}  +
  \delta \Phi^\mu \frac{\overrightarrow{\delta} F}{\delta \Phi^\mu}  \right ) ~,
 \eeq{leftrightderivative}
     we find the corresponding Poisson bracket between the local functional $F$ and $G$
\beq
\{ F, G \} =  \int d\xi d\theta \left ( \frac{ F \overleftarrow{\delta}}{\delta S_\mu} \frac{\overrightarrow{\delta} G}{\delta \Phi^\mu} - \frac{F \overleftarrow{\delta}}{\delta \Phi^\mu} \frac{\overrightarrow{\delta} G}{\delta S_\mu} \right )~.
\eeq{susybracketgeneral} 
 This bracket is super Poisson. In particular we are interested in the local expressions $F (\xi, \theta)$, 
 $$ F(\xi, \theta) = F (\phi, S, D\phi, DS, ...)~,$$
  which are constructed from the basic fields and a finite number of derivatives of those fields.
 This local expression $F(\xi, \theta)$ can be interpreted as
  a functional evaluated at the point $(\xi, \theta)$.  
    These local expressions can be multiplied  and thus they form 
 super Poisson subalgebra of the super Poisson algebra of all functionals on $T^*{\cal L}M$.  
 This super Poisson subalgebra of local expressions is
  the most interesting from the physics point of view.  If we approach this subalgebra formally we end up with the notion of 
   Poisson vertex algebra. 
  The Poisson brackets between the classical generators defined in (\ref{FFFFF1})-(\ref{FFFFF4}) are given by 
  the following expressions:
\ber
\label{bliablia282}  \{{\cal P}(Z),  {\cal P}(Z')\} & = & \left ( 2   \delta(Z,Z') \d'  -  D'\delta(Z,Z') D' + 3  \d' \delta(Z,Z') \right ) {\cal P}(Z') ~, \\
\label{1234abcss}  \{ {\cal P} (Z),  {\cal J}_i (Z') \} & =&  \left ( 2\delta(Z,Z') \d' + 2 \d' \delta(Z,Z')   -  D'\delta(Z,Z') D' \right ) {\cal J}_i(Z')~,\\
 \label{sjdkdkdkdaass} \{{\cal J}_i(Z),  {\cal J}_i(Z')\} &=& -  \delta(Z, Z') {\cal P}(Z') ~,\\
 \label{33902001ssss} \{{\cal J}_1 (Z),  {\cal J}_2(Z')\} & =&  - \delta(Z,Z') {\cal H}(Z') ~,\\
\label{wui9999ss}  \{ {\cal H}(Z),  {\cal J}_1(Z') \} &= &  (2 \delta(Z,Z')\d'- D'\delta(Z, Z') D' + 2 \d' \delta(Z, Z')){\cal J}_2(Z')~,\\
\label{sksks9292ss} \{ {\cal H}(Z), {\cal J}_2(Z')\} & = &  (2 \delta(Z,Z')\d'- D'\delta(Z,Z') D' + 2 \d' \delta(Z,Z')){\cal J}_1(Z')~,\\
\label{22929wiiiwss}    \{ {\cal H}(Z),  {\cal H}(Z')\} & = &
(2\delta(Z,Z') \d' -D'\delta(Z,Z') D' + 3 \d' \delta(Z,Z')){\cal P}(Z')~, \\
 \label{2202wkkwk2h2ss}   \{ {\cal P}(Z),  {\cal H}(Z')\} & =  &   (2\delta(Z,Z') \d'- D'\delta(Z, Z') D' + 3 \d' \delta(Z,Z')){\cal H} (Z') ~.
 \eer{didixkkxshhkaa2}

\end{document}